\documentclass[aps,pre,twocolumn,superscriptaddress]{revtex4}
\usepackage{graphicx,amssymb,amsfonts,amsmath,chemarr,color,commath,braket,hyperref}

\begin{document}

\title{Autologous chemotaxis at high cell density}

\author{Michael Vennettilli}
\thanks{Present address: AMOLF, 1098 XG Amsterdam, The Netherlands}
\affiliation{Department of Physics and Astronomy, University of Pittsburgh, Pittsburgh, Pennsylvania 15260, USA}
\affiliation{Department of Physics and Astronomy, Purdue University, West Lafayette, Indiana 47907, USA}

\author{Louis Gonzalez}
\affiliation{Department of Physics and Astronomy, University of Pittsburgh, Pittsburgh, Pennsylvania 15260, USA}

\author{Nicholas Hilgert}
\thanks{Present address: Department of Systems Biology, Harvard Medical School, Boston, Massachusetts 02115, USA}
\affiliation{Department of Physics and Astronomy, Purdue University, West Lafayette, Indiana 47907, USA}

\author{Andrew Mugler}
\email{andrew.mugler@pitt.edu}
\affiliation{Department of Physics and Astronomy, University of Pittsburgh, Pittsburgh, Pennsylvania 15260, USA}
\affiliation{Department of Physics and Astronomy, Purdue University, West Lafayette, Indiana 47907, USA}

\begin{abstract}
Autologous chemotaxis, in which cells secrete and detect molecules to determine the direction of fluid flow, is thwarted at high cell density because molecules from other cells interfere with a given cell’s signal. Using a minimal model of autologous chemotaxis, we determine the cell density at which sensing fails and find that it agrees with experimental observations of metastatic cancer cells. To understand this agreement, we derive a physical limit to autologous chemotaxis in terms of the cell density, the P\'eclet number, and the length scales of the cell and its environment. Surprisingly, in an environment that is uniformly oversaturated in the signaling molecule, we find that sensing not only can fail, but can be reversed, causing backwards cell motion. Our results get to the heart of the competition between chemical and mechanical cellular sensing and shed light on a sensory strategy employed by cancer cells in dense tumor environments.
\end{abstract}

\maketitle

One of the more remarkable ways that cells detect the flow direction of a surrounding fluid is through a process called autologous chemotaxis \cite{shields2007autologous, fancher2020precision}. In this process, cells secrete a diffusible ligand that they also detect with surface receptors. The flow biases the ligand distribution such that more ligand is detected by receptors downstream than upstream. This imbalance informs the cell of the flow direction. Autologous chemotaxis has been observed for breast cancer \cite{shields2007autologous, polacheck2011interstitial}, melanoma \cite{shields2007autologous}, and glioma cell lines \cite{munson2013interstitial}, as well as endothelial cells \cite{helm2005synergy}.

Tumors and endothelia are, by definition, environments with high cell densities. High cell density poses a challenge to the mechanism of autologous chemotaxis: in addition to detecting its own secreted ligand, a cell will detect ligand secreted by nearby cells. In principle, this exogenous ligand could interfere with the information obtained by a cell from its endogenous ligand. Indeed, it has been shown theoretically that the flow-aligned anisotropy of one cell is reduced by a second cell if both are performing autologous chemotaxis \cite{khair2021two}.

The disruption of autologous chemotaxis at high cell density has been demonstrated experimentally \cite{polacheck2011interstitial}. In fact, at high cell density, cells do not merely stop migrating in the direction of the flow; they migrate against the flow \cite{polacheck2011interstitial}. The reversal is due to a second flow-detection mechanism, distinct from autologous chemotaxis, that relies on focal adhesions and is independent of cell density \cite{polacheck2011interstitial}. Nevertheless, the implication is that for this focal-adhesion-mediated mechanism to dominate at high cell density, autologous chemotaxis must fail at high cell density \cite{polacheck2011interstitial, khair2021two}. Indeed, when the phosphorylation of focal adhesion kinase is blocked, cells at high density once again migrate with the flow, but at a directional accuracy reduced from that of cells at low density without the phosphorylation blocker \cite{polacheck2011interstitial}.

Here we investigate the failure of autologous chemotaxis at high cell density. First, we numerically solve the fluid-flow and advection-diffusion equations for a given cell density in a confined environment, and we find that a cell's anisotropy falls off at a density consistent with experimental observations \cite{polacheck2011interstitial}. To explain this agreement, we derive a physical limit to autologous chemotaxis at high cell density, which successfully predicts the falloff density as a function of the P\'eclet number and the cell and confinement length scales. Finally, we predict that in the presence of an oversaturated amount of background ligand, which can occur, e.g., if some cells secrete but do not absorb ligand, the anisotropy detected by an absorbing cell can actually reverse directions. This reversal is distinct from that due to the focal-adhesion-mediated mechanism \cite{polacheck2011interstitial} but is relevant to scenarios in which non-chemotaxing cells provide additional ligand secretion, as in the case of lymphatic endothelial cells \cite{shields2007autologous}.

\begin{figure}
\includegraphics[width=.75\columnwidth]{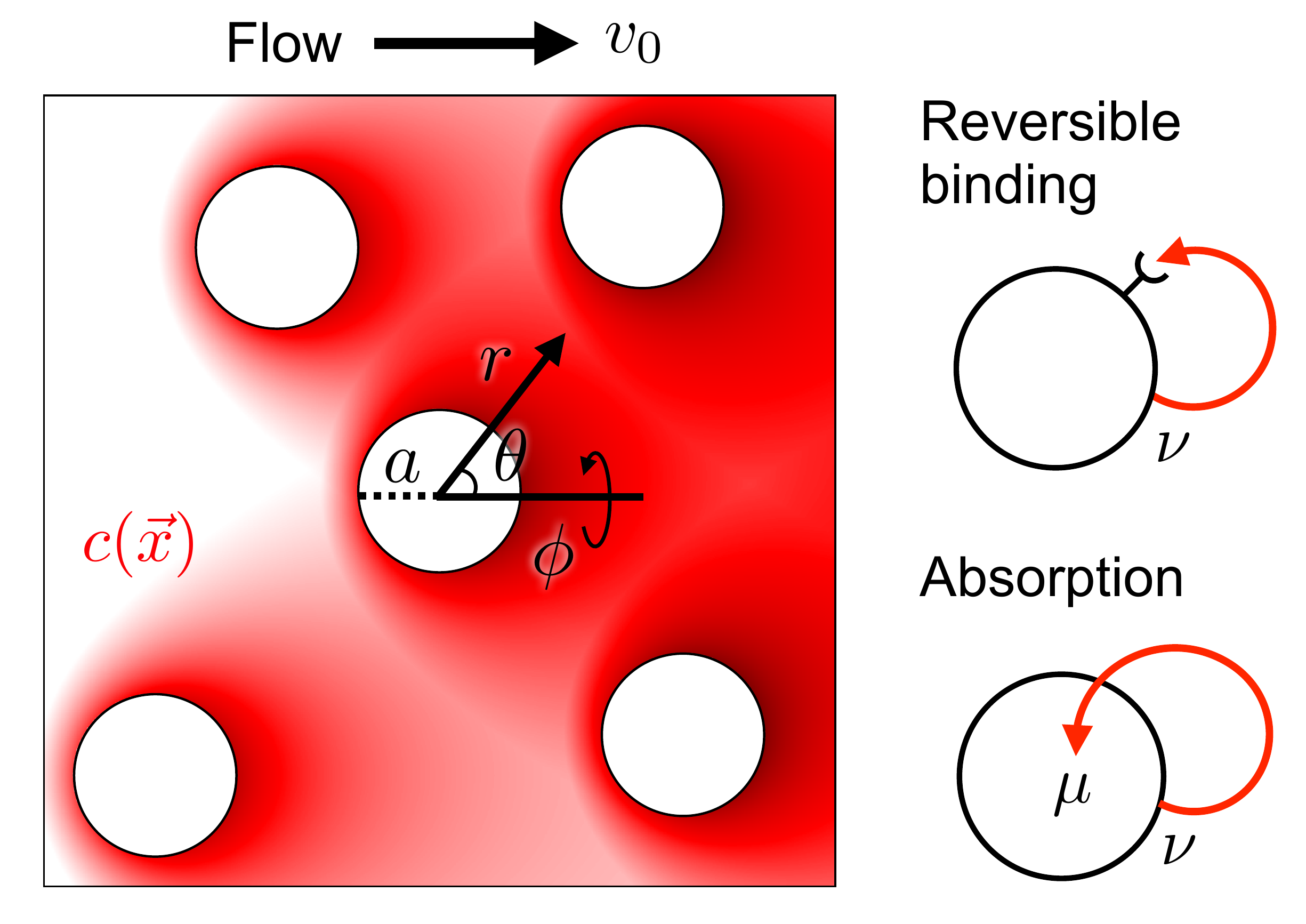}
\caption{Autologous chemotaxis with multiple cells. Each cell of radius $a$ secretes and detects ligand (red). Secretion occurs at rate $\nu$. Detection occurs by reversible binding ($\mu = 0$) or absorption ($\mu > 0$). Fluid flow at background speed $v_0$ biases the ligand concentration $c(\vec{x})$. The anisotropy of the detected signal at the surface of the cell of interest distinguishes its downstream ($\theta = 0$) and upstream ($\theta = \pi$) sides.}
\label{cartoon}
\end{figure}

We consider a minimal model of autologous chemotaxis, in which each cell is a sphere of radius $a$ that is stationary within a background flow of speed $v_0$ (Fig.\ \ref{cartoon}). A cell secretes ligand at a rate $\nu$ and absorbs ligand at a rate $\mu$. We consider two cases for ligand detection: the cell either reversibly binds ligand molecules, corresponding to $\mu = 0$; or the cell absorbs ligand molecules (e.g., via receptor endocytosis), corresponding to a finite value of $\mu$. For absorption, we take $\mu/aD = \pi(\sqrt{17}-1) \approx 10$, where $D$ is the ligand diffusion coefficient, which maximizes sensory precision in this case \cite{fancher2020precision}. In either case, we calculate the normalized anisotropy measure \cite{endres2008accuracy, varennes2017emergent, fancher2020precision}
\begin{equation}
\label{A}
A = \frac{\int d\Omega\ c(a,\theta,\phi) \cos\theta}{\int d\Omega'\ c(a,\theta',\phi')},
\end{equation}
for a given cell from the steady-state ligand concentration $c(r,\theta,\phi)$ at the cell surface $r=a$, where $d\Omega = d\phi d\theta\sin\theta$. The cosine extracts the asymmetry of the concentration between the downstream ($\theta=0$) and upstream ($\theta = \pi$) sides the cell, such that downstream bias corresponds to $A>0$ whereas upstream bias corresponds to $A<0$.

\begin{figure}
\includegraphics[width=\columnwidth]{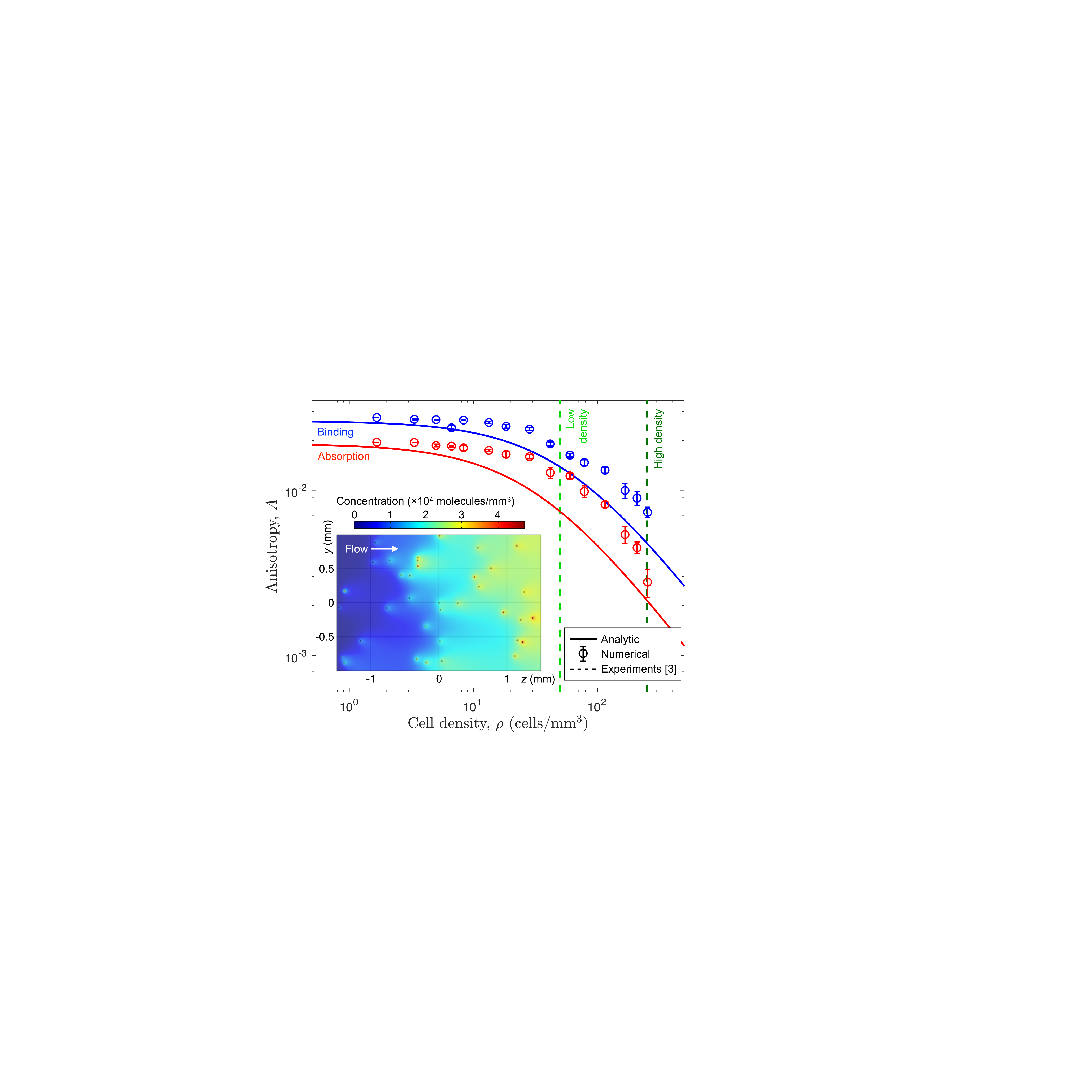}
\caption{Failure of autologous chemotaxis at high cell density. Data points: Anisotropy $A$ at the center cell, averaged from the numerical solution of five configurations per cell density $\rho$ (error bars, standard error). Curves: Analytic approximation (Eq.\ \ref{Arho}). Dashed lines: Experimental seeding densities at which cells were observed to migrate with the flow (low density) and against the flow (high density) \cite{polacheck2011interstitial}. See text for parameter values. Inset: Numerical steady-state concentration profile for an example configuration of 36 cells ($\rho = 60$ cells/mm$^3$) with absorption. }
\label{numerics}
\end{figure}

Although the anisotropy can be obtained analytically to a good approximation for a single cell \cite{fancher2020precision} or two cells \cite{khair2021two}, the presence of multiple cells breaks the symmetry of the flow lines and ligand concentration field, making analytic solution intractable. Therefore we first turn to numerical solution \cite{polacheck2011interstitial}. Specifically, for a given cell configuration, we use finite-element software \cite{code} to solve in steady state (i) the Brinkman equation \cite{brinkman1949calculation} for the flow velocity field, which is appropriate for low-Reynolds-number, low-permeability flow as in the experiments \cite{shields2007autologous, polacheck2011interstitial}; and (ii) the advection-diffusion equation for the ligand concentration, where the solution to the Brinkman equation provides the advection term \cite{supp}. A cell of interest is placed at the center of a box with length $L$, width $W$, and height $H$ (where $L$ is in the flow direction), and other cells are placed randomly such that no cell overlaps a box wall or another cell. An example of the resulting concentration profile is shown in Fig.\ \ref{numerics} (inset). The anisotropy at a given cell density $\rho$ is obtained via Eq.\ \ref{A} from the concentration profile at the center cell's surface, averaged over multiple configurations of the other cells.

We obtain the model parameters from the experiments. A breast cancer (MDA-MB-231) cell is approximately $a = 10\ \mu$m in radius \cite{shields2007autologous, polacheck2011interstitial} and secretes approximately $\nu = 1$ ligand molecule per second \cite{shields2007autologous, fancher2020precision} which diffuses with approximate coefficient $D = 150\ \mu$m$^2$/s \cite{fleury2006autologous}. The cell density experiments \cite{polacheck2011interstitial} were performed with flow velocity $v_0 = 3\ \mu$m/s and permeability $\kappa = 0.1\ \mu$m$^2$ in a chamber with length approximately $L = 3$ mm, width approximately $W = 2$ mm, and height, to the best of our knowledge, on the order of $H = 100\ \mu$m (we will see that our theoretical results do not depend on $H$).

The anisotropy computed from the numerical solution as a function of cell density is shown in Fig.\ \ref{numerics} (data points). Also shown are the experimental densities (dashed lines) at which cells were observed to migrate with the flow (low density) and against the flow (high density) \cite{polacheck2011interstitial}. We see that these densities occur precisely in the regime where the anisotropy transitions from its maximal value to a falloff toward zero. Thus, our numerical solution is quantitatively consistent with the hypothesis that cells at the high density experience a failure of autologous chemotaxis, allowing the focal-adhesion-mediated mechanism to take over \cite{polacheck2011interstitial}.

The numerical solution gives a quantitative prediction for the falloff of anisotropy with cell density, but it does not provide physical intuition. On what parameters does the falloff depend, and why does the transition density occur where it does? To address these questions, we introduce and solve a mean-field model. The result will be a physical limit to autologous chemotaxis that agrees well with the numerical results and predicts the transition density analytically.

\begin{figure}
\includegraphics[width=.75\columnwidth]{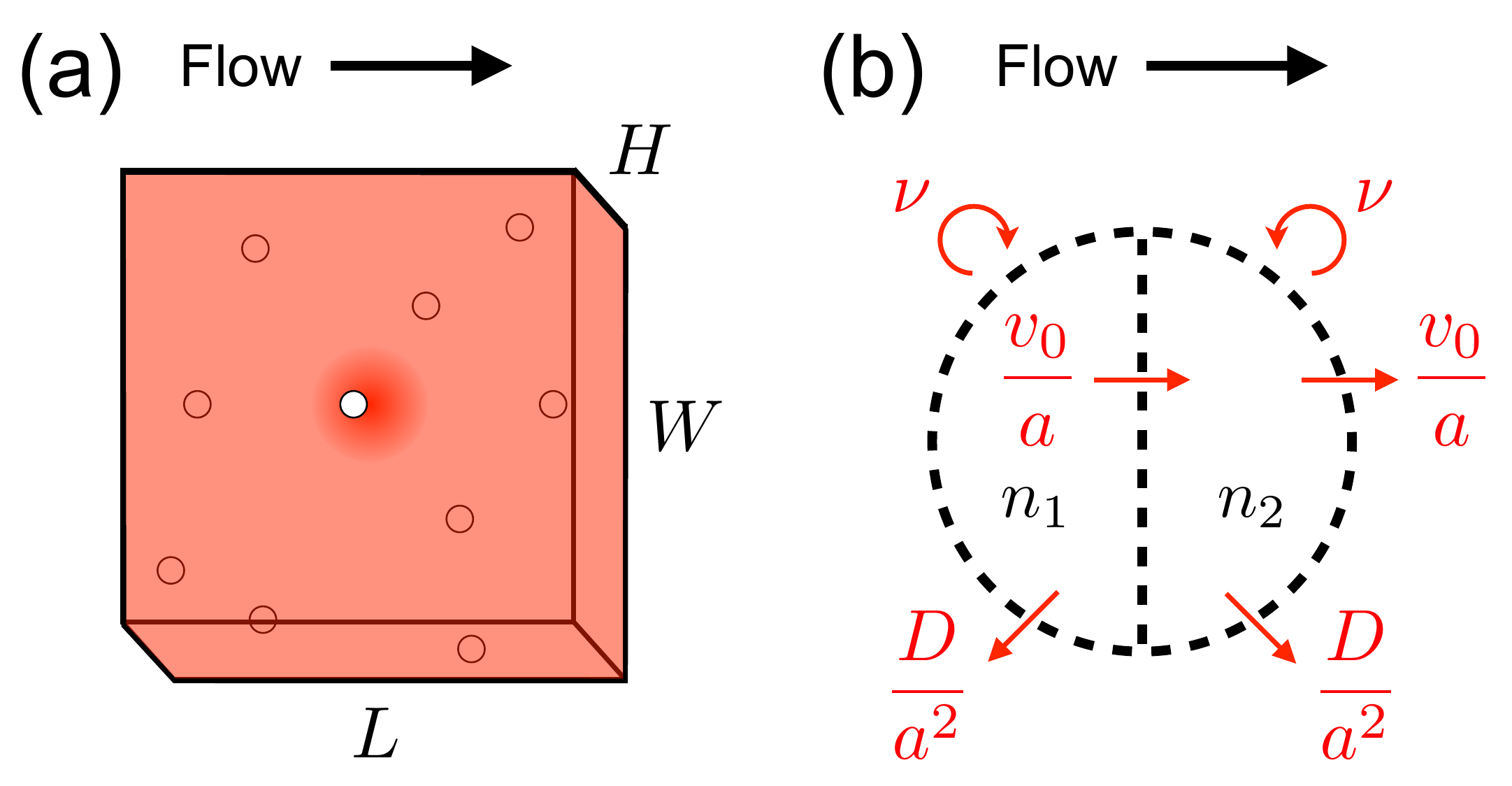}
\caption{Derivation of physical limit. (a) Mean-field model approximates ligand concentration from all cells other than cell of interest as uniform. (b) Anisotropy approximated as difference in ligand number between halves of a permeable cell. Secretion, flow, and diffusion set entry and exit rates.}
\label{limit}
\end{figure}

The mean-field model approximates the ligand from all cells other than the cell of interest as a uniform background concentration $c_0$ [Fig.\ \ref{limit}(a)]. The problem therefore consists of two parts: (i) determine the anisotropy $A$ as a function of $c_0$, and (ii) determine $c_0$ as a function of the cell density $\rho$. For the first part, we will generalize our perturbative solution for a single cell \cite{fancher2020precision} to include a uniform background concentration $c_0$. First, however, we provide a simple argument for how this expression should scale. The result of the scaling argument will turn out to be different from the rigorous solution only by two numerical factors.

Imagine for a moment that the cell is permeable to both the ligand \cite{berg1977physics} and fluid [Fig.\ \ref{limit}(b)]. Permeability to the fluid will neglect the effects of laminar flow around the cell but will permit a simple counting exercise. Specifically, we hypothesize that the anisotropy is roughly equivalent to the difference in the number of molecules in the downstream and upstream halves of the cell, normalized by the total number \cite{mugler2016limits}. These molecule numbers can be obtained approximately by considering the rates of molecule arrival to and departure from each half [Fig.\ \ref{limit}(b)]. Molecules arrive in the vicinity of each half due to secretion, at rate $\nu$. Molecules depart from each half due to diffusion away from the cell, from which a rate can be constructed using the cell radius as $D/a^2$. Finally, molecules depart the upstream half due to flow, at a similarly constructed rate $v_0/a$. Molecules also depart the downstream half, but this departure is compensated by the arrival of the upstream half's molecules, resulting in no net loss due to flow. The molecule number in each half is then the ratio of arrival rates to departure rates, giving a difference of
\begin{equation}
\label{Dn}
\Delta n \sim \frac{\nu}{D/a^2} - \frac{\nu}{D/a^2 + v_0/a} \approx \frac{\epsilon\nu a^2}{D},
\end{equation}
where the second step introduces the P\'eclet number $\epsilon \equiv v_0a/D$ and recognizes it as a small parameter (indeed, the experimental values above give $\epsilon = 0.2$). The molecule number in the whole cell is twice that of either half, or, ignoring the factor of $2$, $n \sim \nu a^2/D$.

Now consider the addition of a uniform background concentration of ligand $c_0$. The background does not change the difference $\Delta n$, but it does change the total molecule number $n$. Specifically, $n$ increases by the number of background ligand molecules in the cell volume, which scales as $c_0a^3$, giving $n \sim \nu a^2/D + c_0a^3$. Combining this expression with Eq.\ \ref{Dn}, we obtain
\begin{equation}
\label{Ac_est}
A \sim \frac{\Delta n}{n} \sim \frac{\epsilon\nu}{\nu + c_0aD}
\end{equation}
as a scaling estimate for the anisotropy.

We compare this estimate with a rigorous calculation of the anisotropy in the presence of a uniform background concentration that we perform by generalizing our previous work \cite{fancher2020precision} using the P\'eclet number as a perturbation parameter \cite{acrivos1962heat, barman1996flow}. The result is \cite{supp}
\begin{equation}
\label{Ac_abs}
A = \frac{f\epsilon(\nu-\mu c_0)/8}{\nu+4\pi c_0aD},
\end{equation}
where $f \equiv (1+\mu/8\pi aD)^{-1}$. In the limit of reversible binding ($\mu\to0$), this result simplifies to
\begin{equation}
\label{Ac_bind}
A = \frac{\epsilon\nu/8}{\nu+4\pi c_0aD}.
\end{equation}
Comparing Eqs.\ \ref{Ac_est} and \ref{Ac_bind}, we see that the scaling estimate is accurate apart from two numerical factors ($1/8$ in the numerator and $4\pi$ in the denominator).

For the second part, determining the background concentration $c_0$ as a function of the cell density $\rho$, we present a simple flux argument. Molecules enter the box due to secretion by cells and leave the box due to (i) absorption by cells and (ii) flow. Specifically, the flux of molecules entering the box per unit time is equal to the secretion rate per cell $\nu$ times the number of cells in the box $\rho LWH$. The flux of molecules leaving the box per unit time contains two terms. The first is equal to the absorption rate at each cell's surface $\mu c_0$ times the number of cells in the box $\rho LWH$, where we have approximated the surface concentration as the background concentration $c_0$. The second is equal to the background concentration $c_0$ times the volume of fluid leaving the box per unit time, which is the flow velocity $v_0$ times the area of the outlet wall $WH$ [Fig.\ \ref{limit}(a)]. At steady state, these fluxes balance,
\begin{equation}
\label{flux}
\underbrace{\nu\rho LWH}_{\rm flux\ in} = \underbrace{\mu c_0\rho LWH + c_0v_0WH}_{\rm flux\ out}.
\end{equation}
Solving Eq.\ \ref{flux} for $c_0$, we obtain
\begin{equation}
\label{c0}
c_0 = \frac{\nu L\rho}{\mu L\rho + v_0},
\end{equation}
which relates the background concentration $c_0$ to the cell density $\rho$.

Inserting Eq.\ \ref{c0} into Eq.\ \ref{Ac_abs}, we obtain
\begin{equation}
\label{Arho}
A = \frac{f\epsilon/8}{1+\rho/\rho_c},
\end{equation}
where $\rho_c \equiv g\epsilon/4\pi a^2L$ with $g \equiv (1+\mu/4\pi aD)^{-1}$. Equation \ref{Arho} demonstrates that the anisotropy falls off with cell density, as expected, and gives the physical parameters on which the falloff depends. In particular, $\rho_c$ is the transition density, where the anisotropy is half of its maximal value. The transition density can never be larger than its value in the reversible binding limit ($\mu\to0$), i.e.,
\begin{equation}
\label{rhoc}
\rho_c \le \frac{\epsilon}{4\pi a^2 L}.
\end{equation}
Equation \ref{rhoc} reveals that the transition density scales with the P\'eclet number $\epsilon$ and inversely with a characteristic volume consisting of the cell surface area $4\pi a^2$ and the system lengthscale in the flow direction $L$. Equation \ref{rhoc} represents a physical limit: the maximum cell density for which autologous chemotaxis can succeed, dependent only on the physical properties of the system $\epsilon$, $a$, and $L$. It also has an intuitive interpretation: scaling lengths by $a$ such that $\rho_ca^3 \le (\epsilon/4\pi)/(L/a)$, we see that the maximum density increases as (i) the P\'eclet number $\epsilon$ increases or (ii) the flow-aligned confinement length $L$ decreases. The former makes sense because flow is easier to detect with diffusible molecules for large P\'eclet number. The latter makes sense because a smaller flow-aligned length places more cells transverse to the flow, where their secreted ligand interferes less with the autologous chemotaxis mechanism of a given cell \cite{khair2021two}.

We see in Fig.\ \ref{numerics} that Eq.\ \ref{Arho} (curves) agrees well with the numerical results (data points). In particular, the theory predicts the $\rho^{-1}$ falloff observed numerically and predicts the observed transition densities in both the reversible binding (blue) and absorbing (red) cases up to a factor of approximately two. The agreement is particularly remarkable because the theory ignores the laminar flow lines from all cells but the cell of interest and assumes that the ligand from these cells is uniformly distributed, which, as seen in the inset of Fig.\ \ref{numerics}, is clearly not the case. We therefore conclude that these details are not essential to the basic physics, and the mean-field theory suffices to quantitatively capture the key dependencies.

Returning now to Eq.\ \ref{Ac_abs}, we see that $A$ becomes negative if $c_0 > c_0^*$, where $c_0^* \equiv \nu/\mu$. This result means that for a sufficiently large background concentration, the bias in the ligand signal detected by the cell reverses direction. This result only occurs for absorption, not reversible binding, because in the binding case ($\mu\to0$) we see that $c_0^* \to\infty$. It also cannot occur if the background concentration is supplied by other similar cells at density $\rho$, as has been assumed so far, at least under the approximation that the background concentration is uniform. Specifically, writing Eq.\ \ref{c0} as $c_0 = c_0^*/(1+v_0/\mu L\rho)$ shows that $c_0 < c_0^*$ in this case, no matter how large $\rho$ becomes. However, the result can occur if the background concentration is supplied independently of the secrete-and-detect mechanism employed by the cell of interest, for example by other cells of a different type. While not relevant to the existing experiments \cite{polacheck2011interstitial} per se, excess exogenous ligand is a highly plausible scenario {\it in vivo}. The particular ligand types involved in autologous chemotaxis (chemokines CCL19 and CCL21) are omnipresent in the tumor microenvironment \cite{korbecki2020cc, lin2014ccl21, sharma2020ccl21, cheng2018ccl19} and are explicitly secreted by lymphatic endothelial cells that coexist with the cancer cells in which autologous chemotaxis was first observed \cite{shields2007autologous}.

Why does the anisotropy reverse sign for large background concentration? The reason is that, because of the flow, the background ligand is absorbed preferentially at the upstream side of the cell (like mist preferentially wetting one side of an object in a steady wind). For a sufficiently large background concentration and absorption rate, this upstream bias can overpower the downstream bias of the autologous chemotaxis mechanism. This is why there is no reversal for the binding case: unlike absorption, binding is an equilibrium process that is unaffected by the symmetry-breaking drift of the uniform ligand field. This is also why there is no reversal if the background concentration is supplied by like cells: sufficient absorption at the cell of interest would imply large absorption by the other cells, which would then necessarily lower the background concentration.

We have investigated the physical mechanisms behind a remarkable form of flow detection, and how the chemical sensing on which this process depends, breaks down at high cell densities. By numerically solving the fluid-flow and advection-diffusion equations, we predicted the regime of cell densities at which sensing fails and found that our predictions agree with experimental observations. By solving a mean-field model and using simple scaling arguments, we revealed how this critical density depends on the system parameters, thereby providing a physical limit for the maximum cell density at which this type of sensing can succeed. Surprisingly, we observed that in an environment super-saturated in the sensed chemical (i.e.\ more than that secreted by the cells themselves), the signal can reverse directions due to an abundance of exogenous chemical absorbed at the upstream side of the cell, predicting a chemically mediated reversal distinct from the mechanically mediated reversal observed in experiments. The results herein deepen our physical understanding of a process that guides cancer cell invasion \cite{shields2007autologous, polacheck2011interstitial, munson2013interstitial}, particularly in the dense and complex environments that characterize tumors \cite{spill2016impact, hinshaw2019tumor}.

At low cell densities, autologous chemotaxis outcompetes pressure sensing at both high and low flow speeds \cite{polacheck2011interstitial}. Our previous work suggests that cells perform autologous chemotaxis with a precision that approaches the physical limit \cite{fancher2020precision}. This finding helps explain why changing the physical environment by increasing cell density is sufficient to cause autologous chemotaxis to fail. On the other hand, the physical limit to the precision of pressure sensing \cite{bouffanais2013physical}, with which autologous chemotaxis competes, is poorly understood for these cells \cite{polacheck2011interstitial}. Therefore, it is an open question whether pressure sensing is subdominant to autologous chemotaxis at low cell density for biological reasons or by physical necessity.

Autologous chemotaxis fails at high cell densities because the concentration profile in the vicinity of a given cell is homogenized due to ligand secreted by other cells. In principle, this homogeneity could be avoided by known mechanisms that preserve spatial or temporal heterogeneity. For example, cells generally are not distributed uniform-randomly as assumed here. In the experiments \cite{polacheck2011interstitial}, cells were observed to migrate in chains that followed the flow lines. In tumors, cells are known to migrate in chains, sheets, or clusters during metastatic invasion \cite{friedl2009collective, cheung2013collective}. Alternatively, secreting and detecting the ligand in temporal pulses could help a given cell distinguish its endogenous ligand from exogenous ligand secreted by other cells at other times. Although the dynamics of chemokine release are still poorly understood, single temporal pulses of chemokine release are sufficient to generate responses at the cell level \cite{sarris2016manipulating}.

Finally, a ubiquitous way that cells at high densities detect weak signals, including concentrations \cite{gregor2007probing, little2013precise, erdmann2009role} and concentration gradients \cite{rosoff2004new, malet2015collective, ellison2016cell}, is by acting collectively \cite{mugler2016limits, camley2016emergent, fancher2017fundamental, varennes2017emergent}. While here we have elucidated how high cell density is detrimental to autologous chemotaxis performed by a single cell, high cell density may be beneficial if an analogous computation is performed at the collective level. Whether autologous chemotaxis benefits from collective effects remains an open question.

\begin{acknowledgments}
We thank Hye-ran Moon and Bumsoo Han for help with COMSOL in the early stages of the work. This work was supported by Simons Foundation Grant No.\ 376198 and National Science Foundation Grant Nos.\ MCB-1936761 and PHY-1945018.
\end{acknowledgments}


\section*{SUPPLEMENTAL MATERIAL}

\subsection{Details of numerical solution}

We obtain the numerical solution using the finite-element software COMSOL. The boundary conditions are set as follows. For the Brinkman equation, cell surfaces and side walls are no-slip boundaries, whereas the upstream wall is an inlet boundary with flow speed $v_0$, and the downstream wall is an outlet boundary with zero pressure. For the advection diffusion equation, the side walls are no-flux boundaries, the upstream wall is an inflow boundary (zero concentration), the downstream wall is an outflow boundary (no flux), and cell surfaces are flux boundaries with secretion and absorption rates $\nu$ and $\mu$, respectively.

Specifically, the flux boundary condition in COMSOL reads
\begin{equation}
\label{boundary}
-\vec{n}\cdot(\vec{J}+\vec{v}c) = k_c(c_b-c).
\end{equation}
Because the normal vector $\vec{n}$ is parallel to the outward flux $\vec{J} = D\partial_r c\ \hat{r}$ at the spherical boundary $r=a$, and the flow velocity $\vec{v}$ vanishes at the cell surface, Eq.\ \ref{boundary} simplifies to
\begin{equation}
-D\partial_r c|_a = k_cc_b - k_cc|_a.
\end{equation}
Comparing this condition with our secretion/absorption condition (Ref.\ [2] of the main text)
\begin{equation}
\label{boundary2}
-4\pi a^2D\partial_r c|_a = \nu - \mu c|_a,
\end{equation}
we see that
\begin{equation}
k_c = \frac{\mu}{4\pi a^2}, \qquad c_b = \frac{\nu}{\mu}.
\end{equation}
We use these expressions, with $a = 10\ \mu$m, $D = 150\ \mu$m$^2$/s, $\nu = 1$/s, and a given $\mu$, to set $k_c$ and $c_b$. For the absorbing case, we use $\mu/aD = \pi(\sqrt{17}-1)$ to set $\mu$. For the binding case, because we cannot take $\mu$ strictly to zero, we use $\mu/aD = 10^{-3} \ll 1$ to set $\mu$.

The permeability estimated in the experiments is $\kappa = 0.1\ \mu$m$^2$ (Ref.\ [3] of the main text), which is much lower than the cross-sectional area of a cell, i.e., $\kappa/a^2 = 10^{-3} \ll 1$. We find that using such a low permeability in COMSOL introduces numerical instability. Therefore we use a value ten times higher, $\kappa = 1\ \mu$m$^2$, for which $\kappa/a^2 = 10^{-2}$, which still satisfies $\kappa/a^2 \ll 1$. We find essentially no difference between $\kappa/a^2 = 10^{-3}$ and $\kappa/a^2 = 10^{-2}$ in the single-cell analytic solution (Ref.\ [2] of the main text) because the system is so deeply within the low-permeability regime.

To correct for any numerical error introduced by COMSOL, we calibrate the numerical anisotropy values to the analytic solution $A_a = \epsilon/(8+\mu/\pi a D)$ for a single cell (Eq.\ 4 of the main text with $c_0 = 0$), which we know is accurate to within $0.4\%$ (Ref.\ [2] of the main text). Specifically, for each of the two $\mu$ values, we calculate the anisotropy $A_n$ from the numerical solution for a single cell in a sufficiently large box to be approximated as infinite. We find that the existing box dimensions $L = 3$ mm and $W = 2$ mm are sufficiently large that $A_n$ no longer changes as either is increased, whereas the box dimension $H = 100\ \mu$m must be increased to $1$ mm to satisfy this condition. We then multiply the numerical anisotropy values (with $H$ back to $100\ \mu$m) by $A_a/A_n$ to enforce the calibration.

COMSOL and Matlab code is freely available (Ref.\ [9] of the main text).

\subsection{Analytic solution with background concentration}

We solve for the anisotropy for a single cell with a uniform ligand background concentration following our previous solution for no background concentration (Ref.\ [2] of the main text). We begin by defining the non-dimensionalized variables
\begin{equation}
\chi \equiv c a^3, \qquad x \equiv \frac{r}{a}
\end{equation}
and parameters
\begin{equation}
\beta \equiv \frac{\nu a^2}{4\pi D}, \qquad \alpha \equiv \frac{\mu}{4\pi aD}, \qquad \zeta \equiv \frac{\sqrt{\kappa}}{a}.
\end{equation}
In these terms, the steady-state diffusion-with-drift equation for the ligand concentration reads
\begin{equation}
\label{dd}
0 = \nabla_x^2 \chi - \epsilon \vec{u}\cdot \vec{\nabla}_x\chi,
\end{equation}
where $\epsilon \equiv v_0a/D$ is the P\'eclet number, and
\begin{equation}
\label{u}
\vec{u}(x,\theta) = \frac{\vec{v}(x,\theta)}{v_0} = u_x(x) \cos\theta\ \hat{x} - u_{\theta}(x) \sin\theta\ \hat{\theta}
\end{equation}
is the non-dimensionalized flow field, with
\begin{equation}
\label{uu}
\begin{gathered}
u_x(x) \equiv 1 -\frac{Z}{x^3} + \frac{3\zeta}{x^2}\left( 1+ \frac{\zeta}{x} \right) e^{(1-x)/\zeta}, \\
u_{\theta}(x) \equiv 1 + \frac{2Z}{x^3} -\frac{3}{2x}\left(1+\frac{\zeta}{x}+\frac{\zeta^2}{x^2} \right) e^{(1-x)/\zeta},
\end{gathered}
\end{equation}
and $Z \equiv 1 + 3\zeta +3 \zeta^2$ (Ref.\ [16] of the main text).

We solve Eq.\ \ref{dd} through the method of matched asymptotic expansions (Ref.\ [15] of the main text), using $\epsilon\ll1$ as our expansion parameter. The inner expansion, valid near the cell surface, is
\begin{equation}
\label{chi}
\chi(x,\theta) = \chi_0(x,\theta) + \epsilon \chi_1(x,\theta) + \mathcal{O}(\epsilon^2).
\end{equation}
We will see that $\chi_0$ is isotropic, and therefore to find the anisotropy, we will need to solve the inner expansion out to $\chi_1$.
The inner expansion satisfies the boundary condition at the cell surface, which reads
\begin{equation}
\label{bc}
-\partial_x\chi|_{x=1} = \beta - \alpha \chi(1,\theta)
\end{equation}
from Eq.\ \ref{boundary2}.

The outer expansion is valid far from the cell surface. Therefore we define a rescaled distance $s \equiv \epsilon x$ and denote the outer expansion as $X(s,\theta)$. In terms of $s$, Eq.\ \ref{dd} reads
\begin{equation}
\label{Xeq}
0 = \nabla_s^2 X - \vec{u}(s/\epsilon,\theta) \cdot \vec{\nabla}_s X.
\end{equation}
From Eqs.\ \ref{u} and \ref{uu}, we see that $\vec{u}(s/\epsilon,\theta)$ has components that scale as
\begin{equation}
\label{use}
u_x\left(\frac{s}{\epsilon}\right) = 1 - \frac{Z\epsilon^3}{s^3}, \qquad
u_\theta\left(\frac{s}{\epsilon}\right) = 1 + \frac{2Z\epsilon^3}{s^3},
\end{equation}
where we neglect the $e^{-s/\epsilon}$ dependence because it falls off faster than any power of $\epsilon$ for $\epsilon\ll1$. Because we only take the inner expansion to first order in $\epsilon$, we neglect the third-order terms in Eq.\ \ref{use}. Therefore, Eq.\ \ref{Xeq} becomes
\begin{equation}
\label{Xeq2}
0=\nabla_s^2 X -\cos\theta \frac{\partial X}{\partial s} +\frac{\sin\theta}{s} \frac{\partial X}{\partial \theta}.
\end{equation}
Because there is no explicit $\epsilon$ dependence in Eq.\ \ref{Xeq2}, we do not define a perturbative expansion for $X$ as we do for $\chi$. Instead, we solve Eq.\ \ref{Xeq2} directly. We will see that the solution is an expansion whose coefficients can depend on $\epsilon$, and this dependence will be determined by the asymptotic matching.
The outer expansion satisfies the boundary condition at infinity,
\begin{equation}
\label{Xbc}
X(s\to\infty,\theta) = c_0a^3,
\end{equation}
which accounts for the uniform background concentration $c_0$.

We perform the asymptotic matching by requiring $\chi$ and $X$ to have the same functional form in some common region where both expansions meet. This is achieved by matching the behavior of $\chi$ as $x\rightarrow \infty$ to that of $X$ as $s\rightarrow 0$.

\subsubsection{Outer expansion}

We write the outer expansion in the form
\begin{equation}
\label{Xsplit}
X(s,\theta) = c_0a^3 + \tilde{X}(s,\theta).
\end{equation}
The first term satisfies the boundary condition in Eq.\ \ref{Xbc} explicitly, and therefore $\tilde{X}$ must vanish as $s\to\infty$. The first term also vanishes in Eq.\ \ref{Xeq2}, and therefore $\tilde{X}$ must solve Eq.\ \ref{Xeq2}. The solution that vanishes at infinity is (Ref.\ [2] of the main text)
\begin{equation}
\label{X0}
\tilde{X}(s,\theta) = \frac{e^{\frac{s}{2}\cos\theta}}{\sqrt{s}} \sum\limits_{\ell = 0}^\infty C_\ell K_{\ell+1/2}(s/2)Y_{\ell}^{0}(\theta),
\end{equation}
where the $Y_\ell^m$ are spherical harmonics, and the $K$ are modified Bessel functions of the second kind. The coefficients $C_\ell$ will be determined by the matching condition.

\subsubsection{Inner expansion}

Inserting Eq.\ \ref{chi} into Eqs.\ \ref{dd} and \ref{bc}, we see that, to zeroth order in $\epsilon$, we have
\begin{equation}
\label{chi0eq}
0 = \nabla_x^2 \chi_0
\end{equation}
with the boundary condition
\begin{equation}
\label{bc0}
-\partial_x \chi_0 |_{x=1} = \beta -\alpha\chi_0(1,\theta).
\end{equation}
The solution to Eq.\ \ref{chi0eq} possessing azimuthal symmetry is
\begin{equation}
\label{chi0full}
\chi_0(x,\theta) = \sum_{\ell=0}^\infty \left( A_{0\ell} x^\ell + \frac{B_{0\ell}}{x^{\ell+1}} \right) Y_{\ell}^0(\theta).
\end{equation}
Inserting Eq.\ \ref{chi0full} into Eq.\ \ref{bc0} and noting that the spherical harmonics are linearly independent, we obtain the system
\begin{equation}
\label{bc0AB}
-\ell A_{0\ell} + (\ell+1)B_{0\ell} = \sqrt{4\pi} \beta\delta_{0\ell} - \alpha(A_{0\ell} +B_{0\ell}),
\end{equation}
where we have used the fact that $Y^0_0 = 1/\sqrt{4\pi}$.
Because the Bessel functions in the outer expansion (Eq.\ \ref{X0}) decrease as a function of $s = \epsilon x$, matching will fail if $\chi_0$ contains terms that increase as a function of $x$. Therefore, we must have $A_{0\ell} = 0$ for $\ell > 0$. Eq.\ \ref{bc0AB} for $\ell > 0$ then reads $(\ell+1)B_{0\ell} = -\alpha B_{0\ell}$, but because $\alpha$ is nonnegative we must also have $B_{0\ell} = 0$ for $\ell>0$. Finally, Eq.\ \ref{bc0AB} for $\ell = 0$ reads $B_{00} = \sqrt{4\pi}\beta-\alpha (A_{00} + B_{00})$, or $B_{00} = (\sqrt{4\pi}\beta-\alpha A_{00})/(1+\alpha)$. Together, these identifications reduce Eq.\ \ref{chi0full} to
\begin{equation}
\label{chi0}
\chi_0(x) = \frac{A_{00}}{\sqrt{4\pi}} + \frac{\gamma}{x},
\end{equation}
where
\begin{equation}
\label{gamma}
\gamma \equiv \frac{\beta-\alpha A_{00}/\sqrt{4\pi}}{1+\alpha}.
\end{equation}
$A_{00}$ will be determined by the matching condition.

To first order in $\epsilon$, we have
\begin{equation}
\label{chi1eq}
0 = \nabla_x^2 \chi_1 - \vec{u}\cdot\vec{\nabla}_x\chi_0
\end{equation}
with the boundary condition
\begin{equation}
\label{bc1}
-\partial_x \chi_1 |_{x=1} = -\alpha\chi_1(1,\theta).
\end{equation}
Note that unlike in the outer expansion which is valid far from the cell, in Eq.\ \ref{chi1eq} we must use the full form of $\vec{u}$ (Eqs.\ \ref{u} and \ref{uu}) because we are near the cell surface. The solution to Eq.\ \ref{chi1eq} possessing azimuthal symmetry is (Ref.\ [2] of the main text)
\begin{align}
\chi_1(x,\theta) =\ &\frac{\gamma}{8}\sqrt{\frac{4\pi}{3}}F(x)Y_1^0(\theta) \nonumber\\
\label{chi1full}
	&+ \gamma\sum_{\ell=0}^\infty \left( A_{1\ell} x^\ell + \frac{B_{1\ell}}{x^{\ell+1}} \right) Y_{\ell}^0(\theta),
\end{align}
where
\begin{align}
F(x) =\ &4 - \frac{4(2\zeta+1)}{x^2} + \frac{2(1+3\zeta+3\zeta^2)}{x^3} \nonumber\\
	&+ \frac{\zeta^2e^{1/\zeta}}{x^3}\left[\left(\frac{x^3}{\zeta^3}-\frac{x^2}{\zeta^2}+\frac{2x}{\zeta}-6\right)e^{-x/\zeta}\right. \nonumber\\
\label{f}
	&\qquad\qquad\quad\left.-\ \frac{x^4E_1(x/\zeta)}{\zeta^4}\right],
\end{align}
and $E_1(y) \equiv \int_1^\infty dt\ e^{-ty}/t$.
Noting that $F'(1) = F(1) = w$, where
\begin{equation}
\label{w}
w \equiv 1 + \frac{1}{\zeta} - \frac{e^{1/\zeta}E_1(1/\zeta)}{\zeta^2},
\end{equation}
we insert Eq.\ \ref{chi1full} into Eq.\ \ref{bc1} to obtain the system
\begin{align}
&\sqrt{\frac{4\pi}{3}}\frac{w}{8}\delta_{1\ell} + \ell A_{1\ell} - (\ell+1)B_{1\ell} \nonumber\\
\label{bc1AB}
&\qquad = \alpha\left(\sqrt{\frac{4\pi}{3}}\frac{w}{8}\delta_{1\ell} + A_{1\ell} + B_{1\ell}\right).
\end{align}
As before, because the Bessel functions in the outer expansion (Eq.\ \ref{X0}) decrease as a function of $s = \epsilon x$, we must have $A_{1\ell} = 0$ for $\ell > 0$. For $\ell > 1$, Eq.\ \ref{bc1AB} then reads $(\ell+1)B_{1\ell} = -\alpha B_{1\ell}$, and again because $\alpha$ is nonnegative we must also have $B_{1\ell} = 0$ for $\ell>1$. Finally, for $\ell = 1$ and $\ell = 0$, Eq.\ \ref{bc1AB} immediately implies $B_{11} = \sqrt{4\pi/3}(w/8)(1-\alpha)/(2+\alpha)$ and $B_{10} = -\alpha A_{10}/(1+\alpha)$, respectively. Together, these identifications reduce Eq.\ \ref{chi1full} to
\begin{align}
\label{chi1}
\chi_1(x,\theta) =\ &\frac{\gamma A_{10}}{\sqrt{4\pi}} \left[ 1 - \frac{\alpha}{(1+\alpha)x} \right] \nonumber\\
	&+ \frac{\gamma}{8}\left[ F(x) + \frac{(1-\alpha)w}{(2+\alpha)x^2}\right]\cos\theta,
\end{align}
where we have recognized that $Y_1^0(\theta) = \sqrt{3/4\pi}\cos\theta$. $A_{10}$ will be determined by the matching condition.

\subsubsection{Matching}
The first two terms in the sum of the outer expansion (Eqs.\ \ref{Xsplit} and \ref{X0}), which will be sufficient for the matching condition, are
\begin{align}
X(s,\theta) = c_0a^3 +\ &\frac{e^{-\frac{s}{2}(1-\cos\theta)}}{2s} \nonumber\\
	&\times\left[C_0 + \sqrt{3}C_1\left(1+\frac{2}{s}\right)\cos\theta + \dots\right],
\end{align}
where we have used $K_{1/2}(s/2) = e^{-s/2}\sqrt{\pi/s}$ and $K_{3/2}(s/2) = e^{-s/2}(1+2/s)\sqrt{\pi/s}$. In the matching limit ($s\to0$) we expand the exponential,
\begin{align}
X(s,\theta) = c_0a^3 +\ &\frac{1}{2s}\left[ 1 - \frac{s}{2}(1-\cos\theta) + \dots \right] \nonumber\\
\label{Xmatch}
	&\times\left[C_0 + \sqrt{3}C_1\left(1+\frac{2}{s}\right)\cos\theta + \dots\right].
\end{align}
For the inner expansion, it will be sufficient in the matching limit ($x\to\infty$) to keep terms out to order $1/x$. To this order, $F(x) = 4$ (Eq.\ \ref{f}), where we have neglected the $e^{-x/\zeta}$ and $E_1(x/\zeta)$ terms because they decay exponentially. Thus, the inner expansion (Eqs.\ \ref{chi}, \ref{chi0}, and \ref{chi1}) becomes
\begin{align}
\chi(x,\theta) =\ &\frac{A_{00}}{\sqrt{4\pi}} + \frac{\gamma}{x} \nonumber\\
\label{chimatch}
	&+ \epsilon\gamma\left\{\frac{A_{10}}{\sqrt{4\pi}} \left[ 1 - \frac{\alpha}{(1+\alpha)x} \right] + \frac{\cos\theta}{2}
	+ \dots\right\}.
\end{align}
To match Eqs.\ \ref{Xmatch} and \ref{chimatch}, we equate like terms in $x$ and $\theta$. The constant terms are
\begin{equation}
\label{m1}
c_0a^3 - \frac{C_0}{4} = \frac{A_{00}}{\sqrt{4\pi}} + \frac{\epsilon\gamma A_{10}}{\sqrt{4\pi}}.
\end{equation}
The terms proportional to $\cos\theta$ are
\begin{equation}
\label{m2}
\frac{C_0}{4} - \frac{\sqrt{3}C_1}{4} = \frac{\epsilon\gamma}{2}.
\end{equation}
Recalling that $s = \epsilon x$, the terms proportional to $1/x$ are
\begin{equation}
\label{m3}
\frac{C_0}{2\epsilon} = \gamma - \frac{\epsilon\gamma A_{10}\alpha}{\sqrt{4\pi}(1+\alpha)}.
\end{equation}
Combining Eqs.\ \ref{m2} and \ref{m3} obtains $C_1 = [\gamma\alpha A_{10}/\sqrt{3\pi}(1+\alpha)]\epsilon^2$. Assuming that $A_{10}$ is order unity (which will be justified post hoc), we see that $C_1$ is second-order in $\epsilon$ and must therefore be neglected because we only took $\chi$ to first order in $\epsilon$. Neglecting $C_1$, Eq.\ \ref{m2} implies
\begin{equation}
C_0 = 2\epsilon\gamma.
\end{equation}
Then, equating terms in Eq.\ \ref{m1} at each order in $\epsilon$ obtains
\begin{equation}
\label{A00}
A_{00} = \sqrt{4\pi}c_0a^3, \qquad A_{10} = -\sqrt{\pi}.
\end{equation}
This completes the matching up to first order in $\epsilon$, which is sufficient to obtain the anisotropy to this order at the cell surface from $\chi$.

\subsubsection{Anisotropy}
In terms of $\chi = ca^3$ and $x = r/a$, the expression for the anisotropy at the cell surface (Eq.\ 1 of the main text) reads
\begin{equation}
A = \frac{\int d\Omega\ \chi(1,\theta) \cos\theta}{\int d\Omega'\ \chi(1,\theta')}.
\end{equation}
In the numerator, any term in $\chi(1,\theta)$ that is constant in $\theta$ will vanish upon integration. Therefore, the numerator will pick out only the term in $\chi_1$ (Eq.\ \ref{chi1}) proportional to $\cos\theta$. In the denominator, the contribution from $\chi_0$ (Eq.\ \ref{chi0}) is non-vanishing. Thus, to leading order in $\epsilon$,
\begin{equation}
A = \frac{\int d\Omega\ \epsilon\chi_1(1,\theta) \cos\theta}{\int d\Omega'\ \chi_0(1)}.
\end{equation}
Inserting the expressions for $\chi_0$ and $\chi_1$ obtains
\begin{equation}
A = \frac{(\epsilon\gamma/8)[ F(1) + w(1-\alpha)/(2+\alpha)]\int d\Omega\ \cos^2\theta}
	{(A_{00}/\sqrt{4\pi} + \gamma)\int d\Omega'}.
\end{equation}
The integrals evaluate to $\int d\Omega\ \cos^2\theta = \int_0^{2\pi}d\phi\int_0^\pi d\theta\sin\theta\cos^2\theta = 4\pi/3$ and $\int d\Omega' = 4\pi$. Recalling that $F(1) = w$, and inserting the expressions for $A_{00}$ (Eq.\ \ref{A00}) and $\gamma$ (Eq.\ \ref{gamma}) and simplifying, we obtain
\begin{equation}
\label{Afinal}
A = \frac{\epsilon w}{8(2+\alpha)}\frac{\beta-\alpha c_0a^3}{\beta+c_0a^3}.
\end{equation}
In the limit of low permeability ($\zeta = \sqrt{\kappa}/a \ll 1$), Eq.\ \ref{w} reduces to $w=2$. With this value, and the original definitions $\beta = \nu a^2/4\pi D$ and $\alpha = \mu/4\pi aD$, Eq.\ \ref{Afinal} becomes Eq.\ 4 of the main text.

\end{document}